\documentclass[onecolumn,epj,final]{svjour}
\usepackage{graphics}
\usepackage{eufrak}

\newcommand{\sgn}{\textrm{sign}}

\usepackage{color}
\usepackage{amsmath}
\usepackage{amssymb}

\def\Xint#1{\mathchoice
   {\XXint\displaystyle\textstyle{#1}}%
   {\XXint\textstyle\scriptstyle{#1}}%
   {\XXint\scriptstyle\scriptscriptstyle{#1}}%
   {\XXint\scriptscriptstyle\scriptscriptstyle{#1}}%
   \!\int}
\def\XXint#1#2#3{{\setbox0=\hbox{$#1{#2#3}{\int}$}
     \vcenter{\hbox{$#2#3$}}\kern-.51\wd0}}

\def\dashint{\Xint-}

\begin{document}
\title{A fractional reaction-diffusion description of supply and demand}
\author{Michael Benzaquen\inst{1}\inst{2} \and Jean-Philippe Bouchaud\inst{2}
}                     
\offprints{}          
\institute{Ladhyx, UMR CNRS 7646, \'Ecole Polytechnique,
91128 Palaiseau Cedex, France \and Capital Fund Management, 23 rue de l'Universit\'e, 75007, Paris, France}
\date{Received: date / Revised version: date}
%
\abstract{We suggest that the broad distribution of time scales in financial markets could be a crucial ingredient to reproduce realistic price dynamics in stylised Agent-Based Models. 
We propose a fractional reaction-diffusion model for the dynamics of {latent} liquidity in financial markets, where agents are very heterogeneous in terms of their characteristic frequencies. Several features of our model are amenable to an exact analytical treatment. We find in particular that the impact is a concave function of the transacted volume (\emph{aka} the ``square-root impact law''), as in the normal diffusion limit. However, the impact kernel decays as $t^{-\beta}$ with $\beta=1/2$ in the diffusive case, which is inconsistent with market efficiency. In the sub-diffusive case the decay exponent $\beta$ takes any value in $[0,1/2]$, and can be tuned to match the empirical value $\beta \approx 1/4$. Numerical simulations confirm our theoretical results. Several extensions of the model are suggested.       
} 
\maketitle
\section{Introduction}
\label{intro}

More than 50 years have passed since Montroll \& Weiss \cite{MontrollWeiss} introduced the continuous-time random walk (CTRW) formalism to account for a broad variety of anomalous diffusion mechanisms. Despite countless achievements in the past decades, non-Gaussian diffusion is still topical in many different fields such as statistical physics, condensed matter physics or biology -- as testified by the present special issue of EPJB. In the present paper, we propose an original application of fractional diffusion to describe the dynamics of supply and demand in financial markets. 

\medskip

In the past few years, the concave nature of the impact of {traded} volume on {asset} prices -- coined the ``square-root impact law'' -- has made its way among the most firmly established stylized facts of modern finance \cite{Grinold,Almgren2005,Toth2011,mastromatteo2014agent,Donier2015,Zarinelli}. 
Several attempts have been made to build theoretical models that account for non-linear market impact, see \emph{e.g.} \cite{gabaix,FGLW}. Following the ideas of T\'oth \emph{et al.} \cite{Toth2011}, the notion of a locally linear ``{latent}'' order-book model (LLOB) was introduced in \cite{MPRL,DonierLLOB}. The latter model builds upon coupled continuous {\it reaction-diffusion equations} for the dynamics of the bid and the ask sides of the latent order book \cite{MPRL} and allows one to compute the price trajectory conditioned to any execution profiles. In the slow execution limit, the LLOB model was shown to match the linear propagator model that relates past order flow to price changes through a power law decaying kernel. \medskip

The propagator model was initially introduced in  \cite{bouchaud2004fluctuations} to solve the so-called diffusivity puzzle: prices are approximately diffusive when the order flow is highly persistent \cite{bouchaud2004fluctuations,Lillo2004,bouchaud2008markets}. In particular, the sign of the order-flow is characterised by an autocorrelation function $C(t)$ that decays as a power law $t^{-\gamma}$ with an exponent $\gamma<1$, defining a {\it long-memory process}. Typically $\gamma$ is found to be $\approx 0.5$ for individual stocks. Naively, the impact of these correlated orders should lead to a super-diffusive price dynamics, with a Hurst exponent $H=1-\gamma/2 > 1/2$. In order to compensate for order flow correlation and restore price diffusivity, the impact of each order, described by a certain kernel or ``propagator'', must itself decay as a power law of time, with an exponent $\beta=(1-\gamma)/2$ \cite{bouchaud2004fluctuations,bouchaud2008markets}. \medskip

One major issue of the original LLOB framework is that its kernel decays with exponent $\beta=1/2$, which cannot fulfil the above relation since $\gamma$ must obviously be positive. Impact relaxation is too quick for $\beta=1/2$, and the price dynamics generated by the LLOB model exhibits significant {\it mean reversion} on short to medium time scales, a feature that is not observed in empirical data. However, the current version  of the LLOB model postulates that market participants are homogeneous, in the sense that the distribution of volumes, reaction times, pricing updates, {etc}. are all {\it thin tailed}. For example, the cancellation of orders is assumed to be a Poisson process with a single cancellation rate $\nu$. \medskip

In reality, different actors in financial markets are known to be highly heterogeneous with very widely distributed volumes and time scales, from High Frequency Traders (HFT) to large institutional investors. This observation suggests various generalisations of the LLOB. The path that we follow in this paper is that of a broad distribution of time scales for agents' intentions and re-evaluation, that naturally leads to a {\it fractional} reaction-diffusion process. Other possible generalisations are discussed in the conclusions. \medskip

The outline of the paper is as follows. We first present the general fractional diffusion framework with death and sources. We then present the fractional latent order-book model (FLOB) and derive its equilibrium shape, in the case of a balanced flow of buy/sell orders. We find that the characteristic V-shape of the latent order book is preserved, leading to a concave impact law. We also show that the corresponding propagator is now described by a tunable exponent $\beta \in [0,1/2]$ that resolves the above mentioned diffusivity puzzle. We finally confront these results to numerical simulations.\medskip

\section{Fractional diffusion with death and sources}
\label{sec:2}

We here present the CTRW framework and adapt it to the question of interest in this paper. The CTRW model describes random walkers that pause for a certain waiting time before resuming their motion. When the average waiting time is finite, the long-time, large scale description of an ensemble of such walkers is the standard diffusion equation. When the average waiting time diverges, the corresponding dynamics is described by the {\it fractional diffusion equation} \cite{Metzler2,Metzler}. \\

Assuming that jump lengths and waiting times are independent, let $\Psi(t)$ denote the waiting time distribution function and $\Lambda(x)$ denote the jump length distribution function. The evolution equation for the density of walkers $\phi(x,t)$ at position $x$ and time $t$ reads \cite{Henry}:
\begin{eqnarray}
\phi(x,t)&=& \Phi(t) \phi(x,0) + \hspace{-0.1cm} \int \hspace{-0.1cm}\textrm dx' \hspace{-0.15cm} \int_{[0,t]} \hspace{-0.3cm} \textrm d t' \hspace{-0.05cm}  \Lambda(x-x') \Psi(t') \phi(x',t-t')\nonumber  \\
&& \quad +\int_{[0,t]} \hspace{-0.3cm} \textrm d t'  \Phi(t')s(x,t-t') \ ,
\label{Ev_eq}
\end{eqnarray}
where $\Phi(t)=1- \int_0^{t} \textrm d t' \Psi(t')$ denotes the often called survival probability function, and where $s(x,t)$ is a general rate-source term allowing for the injection or removal of walkers {(see Appendix A)}. Taking the Fourier-Laplace transform $\mathcal {(FL)}$ of Eq.~\eqref{Ev_eq} yields:
\begin{eqnarray}
\phi(k,p)&=& \Phi(p) \phi_0(k) +  \Lambda(k) \Psi(p) \phi(k,p) + \Phi(p)s(k,p) \ ,  \ \quad   
\label{Ev_eq_FL}
\end{eqnarray}
where $\phi_0(k)\  := \ \phi(k,t=0)$, and where $p \hspace{0.03cm}\Phi(p)=1-\Psi(p)$. We assume the jump lengths to have a zero mean and a finite second moment. In addition, considering that when its time comes a random walker can either resume its motion or disappear with some small probability {  $\eta$}, we allow the distribution of jump lengths to be be non-normalized ($\int \textrm dx \Lambda (x) = 1 - { \eta}$). In the diffusion limit, this is:
\begin{eqnarray}
\Lambda(k)&\approx &  1 - \sigma^2 k^2 - { \eta} \ ,
\label{jl_pdf}
\end{eqnarray}
where $\sigma$ denotes the root mean square of jump lengths.
We assume waiting times to be distributed according to a truncated\footnote{For seasoned readers, truncating the distribution of waiting times may seem unusual in the context of fractional diffusion but is here an essential 
ingredient for describing a stationary order book.} power-law function with tail exponent $\alpha < 1$ and cutoff $t_c = \epsilon^{-1}$, such that (see Appendix B):
\begin{eqnarray}
\Psi(p)&\approx& 1- \tau^\alpha [ (p+\epsilon)^\alpha - \epsilon^\alpha  ]\ ,
\label{wt_pdf}
\end{eqnarray}
where $\tau$ denotes the scale of waiting times. Note that for short times $t \ll t_c$ (equivalently $p \gg \epsilon$) one has: $\Psi(p) \approx 1-\tau^\alpha p^\alpha \approx \exp[-\tau^{ \alpha} p^\alpha]$ consistent with fractional diffusion (see \emph{e.g.} \cite{Henry}), while for long times $t\gg t_c$ (equivalently $p\ll \epsilon$) $\Psi(p) \approx 1-\tilde \tau p$ where $\tilde \tau = \alpha \epsilon ^{\alpha-1} \tau^\alpha$ is the average time between jumps, consistent with normal diffusion.\\

Injecting Eqs.~\eqref{jl_pdf} and \eqref{wt_pdf} into Eq.~\eqref{Ev_eq_FL} and rearranging terms yields:
\begin{equation}
\phi(k,p) = G_{\alpha,\epsilon}(k,p) \big[\phi_0(k) +s(k,p) \big] \ ,
\label{solfourier}
\end{equation}
with:
\begin{eqnarray}
G_{\alpha,\epsilon}(k,p) = \left[ p+ \frac{\omega \,p (k^2+\varphi)}{(p+\epsilon)^\alpha - \epsilon^\alpha} \right]^{-1}  ,
\end{eqnarray}
where $\omega\  :=\ { \sigma^2}/\tau^\alpha$ { and $\varphi \ \hat= \ \eta/\sigma^2$}. Equation \eqref{solfourier} is central as its inverse Fourier-Laplace transform allows to compute the evolution of the walkers density $\phi(x,t)$ for given initial condition and source terms. Unfortunately the inverse Fourier-Laplace transform of the kernel $G_{\alpha,\epsilon}$ is not analytical in the general case. In the following we thus consider the limit cases of short and long times (see Appendix C for a presentation of the problem in terms of partial-differential and integro-differential equations).\\

For $t\ll t_c$, we may note $G_{\alpha,\epsilon}^-(k,p)=p^{\alpha-1}[\,p^\alpha+\omega  (k^2+\varphi) ]^{-1}$. Taking the inverse Fourier-Laplace transform,
one obtains that the kernel in real space is given by the inverse Fourier transform of the Mittag-Leffler\footnote{The Mittag-Leffler function $E_\alpha(z)=F_\alpha(-z)$ is a special function defined by the following series as: $F_{\alpha}(z) = \sum_{j=0}^{\infty}  \frac{(-z)^j}{\Gamma[1+j\alpha]}$.  } function (see \emph{e.g.} \cite{Metzler,Schneider}), $\mathcal G_{\alpha,\epsilon}^-(x,t)=\mathcal F^{-1}\{ F_{\alpha}[\omega (k^2+\varphi)t^{\alpha}] \}$. Note that in the limit $\varphi \rightarrow 0$ the kernel $\mathcal G_{\alpha,\epsilon}^-(x,t)$ can be conveniently written as:
\begin{eqnarray}
\mathcal G_{\alpha,\epsilon}^-(x,t)&=&\frac{1}{\sqrt{4\pi \omega t^{\alpha}}} \, g_{\alpha}\left(\frac{x^2}{4\omega t^\alpha}\right), \quad (\varphi \to 0),
\label{Gg}
\end{eqnarray}
where we introduced the function $g_\alpha$, the shape of which is discussed in \emph{e.g.} \cite{Schneider,Metzler2}. 
For $t\gg t_c$, one obtains $G_{\alpha,\epsilon}^+(k,p)=[\,p+ \tilde \omega(k^2+\varphi) ]^{-1}$ where $\tilde \omega \ \hat= \ { \sigma^2}/\tilde \tau$. Taking the inverse Fourier-Laplace transform yields the normal diffusion kernel (with death): 
\begin{eqnarray}
\mathcal G_{\alpha,\epsilon}^+(x,t)&=&\frac{e^{-\nu t}}{\textstyle \sqrt{4\pi \tilde \omega t}} \, \exp\left(-\frac{x^2 }{4 \tilde \omega  t}\right)  ,
\label{G_norm}
\end{eqnarray}\\
where $\nu = \tilde \omega \varphi$ is the rate of death per unit time.
In both  cases, the general solution in real space can thus be written as the sum of a convolution death/diffusion term and a source contribution:
\begin{eqnarray}
\phi^{\pm}(x,t) &=& [\mathcal{G}_{\alpha,\epsilon}^{\pm}\hspace{-0.05cm}*\phi_0](x,t) +  \mathcal {(FL)}^{-1} \hspace{-0.05cm}\left\{ \mathcal{G}_{\alpha,\epsilon}^{\pm}(k,p)s(k,p)  \right\} . \ \ \ \
\label{gen_sol}
\end{eqnarray}

\section{Fractional latent order-book model}
\label{sec:3}


Following the assumptions of Donier \emph{et al.} \cite{DonierLLOB,DonierWalras}, we posit that the dynamics of the intentions of market participants\footnote{{ Note that given the very weak revealed instantaneous liquidity in the order book, intentions in the sense of latent orders are an essential (and very sound) ingredient in microstructure modelling.  Latent order book models have allowed to to account for important stylised facts such as the square root impact law (see \emph{e.g.} \cite{mastromatteo2014agent,DonierLLOB}).}\smallskip} results from order cancellation or reassessment of their reservation price.  { Such intentions (which are commonly called \textit{latent orders}) materialise into revealed orders in the vicinity the transaction price. The model is zero-intelligence in the sense that agents are heterogeneous and their reservation prices are updated at random}. The crucial difference with the LLOB model of Donier \emph{et al.} \cite{DonierLLOB} is that we now assume that such events occur after a fat-tailed waiting time.\footnote{Let us stress that price reassessments themselves are not fat-tailed distributed, \emph{i.e.} intentions do not follow Levy flights. While such an extension would be interesting, it is not the point of the present paper and is left for future work.\smallskip} The distribution of these waiting times is assumed to decay with a power-law exponent $\alpha < 1$ some until an exponential cutoff at $t_c=\epsilon^{-1}$ (see Appendix B). Note that while such a cutoff is indeed quite realistic (nobody is expected to hold a position forever), it is also indispensable to ensure that the system does not age, which would prevent dynamic stationary states of the latent order book.  \medskip

An important addition to the fractional diffusion equation described in the previous section is the {\it reaction} mechanism, that corresponds to transactions between buy and sell orders that remove volume from the latent order book and set the transaction price. Within this framework, the density of buy $\phi_{\mathrm{b}}(x,t)$  and sell $\phi_{\mathrm{s}}(x,t)$ intentions at price $x$ and time $t$ solve the set of coupled evolution equations in the reference frame of the ``consensus'' price\footnote{We here substantially simplify the discussion given in \cite{DonierLLOB} where $x_t$ in fact follows some additional exogenous dynamics, reflecting the evolution of the agents' expectations about the ``consensus price''. See \cite{DonierWalras} for an extended discussion.}, as given by Eq.~\eqref{Ev_eq} with:
\begin{eqnarray}
s_{\mathrm{b}}(x,t)&=&\lambda \Theta(x_t-x) - R_\mathrm{sb}(x)\\
s_{\mathrm{s}}(x,t)&=&\lambda \Theta(x-x_t) - R_\mathrm{sb}(x) \ ,
\label{}
\end{eqnarray}
where $\Theta$ denotes the Heaviside step function, and where $R_\mathrm{sb}(x)$ describes a reaction rate that instantaneously removes buy and sell ``particles'' as soon as they meet (see~\cite{Henry,FRD3,FRD2} for some work on  fractional reaction-diffusion). Note that transactions remove exactly the same volume of buy and sell orders, justifying the fact that the same rate $R_\mathrm{sb}(x)$ appears in the two equations above.  The term proportional to $\lambda$ corresponds to an incoming flux of buy/sell intentions to the left/right of the transaction price $x_t$. \medskip

The non-linearity arising from the reaction term in the above equations can be abstracted by defining the combination $\psi(x,t) = \phi_{\mathrm{b}}(x,t)-\phi_{\mathrm{s}}(x,t)$, which precisely solves Eq.~(\ref{Ev_eq}) with:
\begin{eqnarray}
s(x,t)&=& \lambda \,\sgn(x_t-x) \, ,
\label{src_depos}
\end{eqnarray}
and where the transaction price $x_t$ is fixed by the condition: 
\begin{eqnarray}
\psi(x_t,t) &=& 0 \, .
\label{price_eq_gen}
\end{eqnarray}
The stationary order-book centred at $x_{\infty}=0$ can be computed from Eqs.~(\ref{gen_sol}) and (\ref{src_depos}) as $\psi_{\mathrm{eq}}(x) = \lim_{t\rightarrow \infty} \psi(x,t)$.  Making use of Eqs.~\eqref{gen_sol}, \eqref{G_norm} and \eqref{src_depos}, one obtains  (see \cite{DonierLLOB}): $\psi_{\mathrm{eq}}(x)=- ({\lambda}/{\nu}) \,\sgn(x)  [1-\exp(-\sqrt{\varphi}|x|)] $.
In the vicinity of the transaction price, the stationary order book can be shown to be locally linear and its local shape is given by:
\begin{eqnarray}
\psi_{\mathrm{eq}}(x) &=& -\mathcal L x +O(x^2)\, ,
\label{linbook}
\end{eqnarray}
where $\mathcal L \ := \  { \lambda \sqrt{\varphi}/\nu} = {\lambda \tilde \tau}/({{{\sigma^2}}\sqrt{\varphi}})$.

\section{Market impact}
\label{sec:4}

In this section we compute and analyze the impact of a meta-order with execution horizon $T$ on the transaction price
Following Donier \emph{et al.}, we introduce the meta-order of volume $Q$ as an extra order flow that falls exactly at the transaction price such that the source term  becomes: 
\begin{eqnarray}
s(x,t)&=& \lambda \,\sgn(x_t-x) + m_t\delta(x-x_t) \ ,
\label{meta_source}
\end{eqnarray}
where $m_t$ denotes the (possibly time dependent) execution rate, with $\int_0^T {\rm d}t \,m_t=Q$. We set $t=0$ to a situation in which the stationary order is well established and focus on the regime where $t<T\ll t_c$.\footnote{Note that in the limit $T\gg t_c$, one recorvers the normal diffusion results of Donier \emph{et al.} \cite{DonierLLOB}} The general solution as given by Eq.~(\ref{gen_sol}) with the source term of Eq.~(\ref{meta_source}) reads:
\begin{eqnarray}
\psi(x,t) &=& [\mathcal{G}_{\alpha,\epsilon}^-*\psi_0](x,t) +  \int_0^t {\rm d}u \, m_{{u}} \mathcal{G}_{\alpha,\epsilon}^-(x-x_{{u}},t-{u})   \quad \nonumber \\
+\,\frac{i\lambda}{\pi}\dashint&\displaystyle   \frac{{\rm d}k}k & \int_0^{\infty}{\rm d}u \, F_{\alpha}\big[\omega(k^2+\varphi)(t-{u})^{\alpha}\big] e^{ik(x-x_{{u}})}\, , \quad
\label{gen_sol_meta_full}
\end{eqnarray}
where $\dashint$ denotes Cauchy's principal value. \medskip

In the following we ``zoom'' into the linear region of the book, close to the transaction price. More precisely, we consider the limit $\varphi, \lambda \rightarrow 0$, while keeping $\mathcal L$ constant. In this limit -- and starting from the equilibrium book $\psi_0(x) = \psi_{\mathrm{eq}}(x)$ -- Eq.~(\ref{gen_sol_meta_full})  becomes: 
\begin{equation}
\psi(x,t) = -\mathcal L x +  \int_0^t {\rm d}u  \frac{m_{{u}}}{\sqrt{4\pi \omega(t-{u})^\alpha}} g_\alpha\left[\frac{(x-x_{{u}})^2}{4\omega(t-{u})^\alpha}\right]    . \quad \quad
\label{gen_sol_meta}
\end{equation}
Making use of Eq.~(\ref{price_eq_gen}) yields the following self-consistent integral equation for the transaction price:
\begin{eqnarray}
x_t &=& \frac1{\mathcal L}  \int_0^t {\rm d}u  \frac{m_{{u}}}{\sqrt{4\pi \omega(t-{u})^\alpha}} g_\alpha\left[\frac{(x_t-x_{{u}})^2}{4\omega(t-{u})^\alpha}\right]    . \quad \quad
\label{price_eq}
\end{eqnarray}
Provided that impact is small (or equivalently in the limit of small execution rates) one has ${(x_t-x_{{u}})^2}\ll {4\omega(t-{u})^\alpha}$, which recovers the propagator limit where the transaction price 
is linearly related to the order flow through a power-law decaying kernel:\footnote{Here, we have made use of the Taylor expansion of the function $g_{\alpha}(y)$ when $y\rightarrow 0$ (see \cite{Schneider,Metzler2}).}
\begin{eqnarray}
x_t &=& \frac{\sqrt{\pi}}{\mathcal L \Gamma[1-\alpha/2]}  \int_0^t {\rm d}u  \frac{m_{{u}}}{\sqrt{4\pi \omega(t-{u})^\alpha}} \ , 
\label{price_eq_slow}
\end{eqnarray}
allowing us to identify the propagator decay exponent $\beta$ with $\min(1/2,\alpha/2)$. Note that for $\alpha<1$ the equality $\beta=(1-\gamma)/2$ can be achieved by the choice 
$\alpha=1-\gamma \in [0,1]$. Hence, the FLOB allows the price to be diffusive at all times in the presence of a persistent order flow. As mentioned in the introduction, real data suggests $\gamma\approx 0.5$ which implies $\alpha\approx 0.5$. 
For a constant execution rate $m_t = m_0 = Q/T$ 
and denoting $I_Q=x_T-x_0$ the impact of a meta-order of size $Q$, one obtains:
\begin{eqnarray}
I_Q &=&  \frac{Q^{1-\alpha/2}}{\mathcal L}\frac{m_0^{\alpha/2}}{(2-\alpha)\Gamma[1-\alpha/2]\sqrt{\omega  }}\ .
\label{imact_small}
\end{eqnarray}
As one can see, Eq.~(\ref{imact_small}) leads to $I_Q \sim Q^{0.75}$ for $\alpha=0.5$, intermediate between a square-root and a linear behaviour. The pure square-root for small execution rates is only recovered in the limit $\alpha=1$ considered by Donier \emph{et al.} \cite{DonierLLOB}. \medskip
 
In the opposite limit however of fast execution  -- more precisely when ${(x_t-x_{{u}})^2}\gg {4\omega(t-{u})^\alpha}$ (but still in the regime $t \ll t_c$) -- one can show that the impact is again given by a square root law  { (see Appendix D):
\begin{eqnarray}
I_Q &=&  h(\alpha)\sqrt{\frac{2Q}{\mathcal L}}    \ ,   \label{largeexecimp}
\end{eqnarray}
where $h(\alpha)=(2-\alpha)^{-1/2} \leq 1$.} Interestingly, the result is smaller than what a purely geometric argument would suggest, where the volume initially contained between $x_0$ and 
$x_T=x_0+I_Q$ is executed against the incoming metaorder. In the latter case one can write:
\begin{eqnarray}
Q &=& \int_0^{I_Q} {\rm d}x \,{\mathcal L }x \quad \Rightarrow \quad I_Q = \sqrt{\frac{2Q}{\mathcal L}}\ ,
\end{eqnarray}
valid for $\alpha \geq 1$. When $\alpha < 1$, liquidity initially outside the interval $[x_0,x_T]$ manages to move inside that interval and meet the incoming metaorder, even in the fast execution limit. This provides more resistance to the metaorder, \emph{i.e.} a slightly smaller impact. 
\medskip

The FLOB thus provides a framework in which concave impact is compatible with persistent order flow, although one expects a cross-over from a $Q^{1-\alpha/2}$ behaviour for ``slow'' execution to a $\sqrt{Q}$ behaviour for ``fast'' execution.\footnote{On the definition of ``slow" and ``fast" for real markets where HFT significantly contribute, see {\cite{ustocome}}.}\medskip

The impact decay after the meta-order execution $t_c \gg t > T$\footnote{For $t\gg t_c$ one eventually recovers the $1/\sqrt{t}$ impact decay predicted by Donier \emph{et al.} \cite{DonierLLOB} in the diffusive limit.} can be computed by replacing the upper boundary of the integrals in Eqs.~(\ref{price_eq}) and (\ref{price_eq_slow}) by $T$. In the limit of small execution rates, one easily  obtains:
\begin{eqnarray}
\frac{I_Q(t>T)}{I_Q}&= &\left(\frac tT\right)^{1-\alpha/2} - \left(\frac {t-T}T\right)^{1-\alpha/2}   \ ,  
\end{eqnarray}
which decays with an infinite slope for $t \to T^+$ and asymptotically { equals $ (1-\alpha/2)(t/T)^{-\alpha/2}$.  In the limit of high execution rates, the impact decay for $t_c\gg t\gg T$ can be conveniently computed as in the slow execution limit and reads:
\begin{equation}
\frac{I_Q(t\gg T)}{I_Q}=  \frac{m_0^{\alpha/2} Q^{(1-\alpha)/2} }{\sqrt{2(2-\alpha)\omega  \mathcal L_\alpha}} \frac{1-\alpha/2}{\Gamma[1-\alpha/2]} \left(\frac tT\right)^{-\alpha/2}  \ ,  
\end{equation}
which only differs from  the low execution result by the prefactor. For $t \to T^+$ the calculation 
is more subtle and relies on the source-free relaxation of the linearised order book profile in the vicinity of the price at the end of the metaorder execution (see \cite{DonierLLOB}). One obtains:
\begin{eqnarray}
\frac{I_Q(t\rightarrow T^+)}{I_Q}&= &1 - \frac {\sqrt{\omega(t-T)^\alpha}}{I_Q} z^*  \ ,  
\end{eqnarray}
where $z^*$ solves: $a^- z - (a^+-a^-)\int_z^{\infty}{\rm d}u\,g_\alpha({u^2}/4) ({u-z})/{\sqrt{4\pi}}  = 0 $ 
with $a^{\pm}=\lim_{ \epsilon \rightarrow  0 } \partial_x\psi(x_T\pm \epsilon,T) $.
    }

\section{Numerical simulation}
\label{sec:5}

\begin{figure}[t!]
\begin{center}
\resizebox{0.99\columnwidth}{!}{%
  \includegraphics{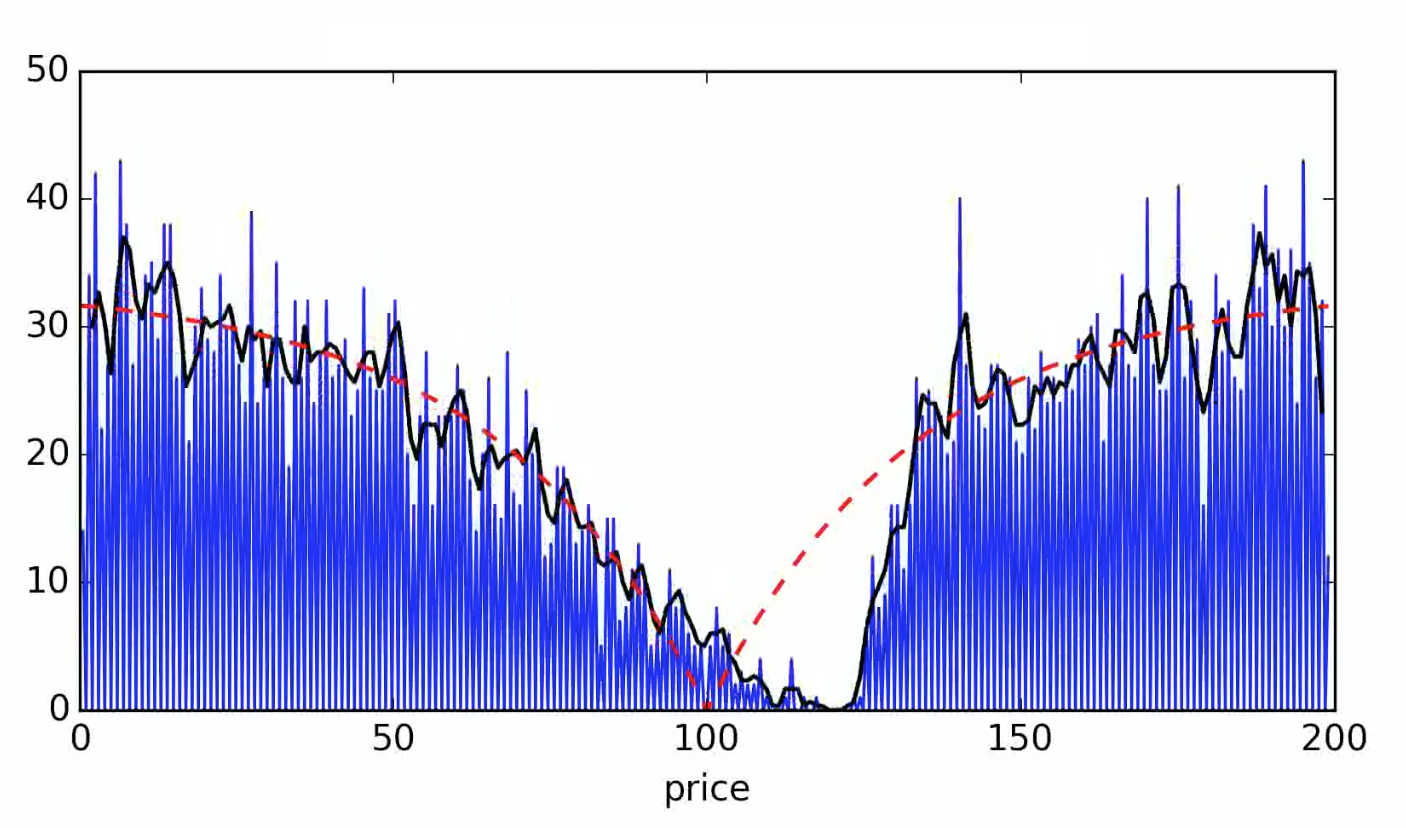}
}
\end{center}
\caption{Numerical simulation of the latent order book during a constant rate buy meta-order execution. The red dashed line corresponds to the equilibrium order book from which the simulation started at $t=0$.}
\label{SimOB}       
\end{figure}

In order to bolster our analytical results we performed a numerical simulation of the model. Because of the peculiar nature of fractional diffusion, such a simulation is more time consuming than a regular reaction-diffusion simulation. This is because time needs be continuous and each particle (order intention) must be treated independently.   Each particle is labeled as \emph{buy}/\emph{sell}, and most importantly \emph{next event time} drawn from a fat-tailed probability distribution function. In order to speed up the simulation we use a heap queue algorithm\footnote{Also called \emph{priority queue algorithm}, the heap queue algorithm is a binary tree in which each parent node stores a value smaller than or equal to its children's, such that in our case the next event is always at the root. New events are pushed into the tree from the bottom and find their place in $\log N$ time {\cite{Heapqueue}}.} to efficiently sort the up-coming events in time. The nature of the next event (diffusion or death) is drawn from a biased distribution depending on the diffusion and cancellation rates. Simultaneously a Poissonian rain of particles 
falls into the book, each particle being naturally labeled with \emph{buy}/\emph{sell} depending on the side of the book it falls in. For the sake of simplicity the spread region between the best buy (bid) and the best sell (ask) is rain-free. For each potentially price changing event, namely each event involving a particle at the bid or ask,  the heap is broken such that the relevant variables -- volume profile, best bid and ask, spread, mid-price, event-time and deposition/cancellation/reaction rates - can be updated and stored. When a buy and a sell ``particle'' find themselves at the same spot, there are immediately removed from the book (reaction). \medskip

The simulation is initialized from the theoretical equilibrium situation.
Reflective boundary conditions are implemented to ensure flat slopes at the (far away) edges of the book.
A meta-order execution can be implemented through an additional rain of buy (resp. sell) particles that fall precisely at the best ask (resp. bid) with given rate $m_t$ (see Fig.~\ref{SimOB} for an illustration of a typical numerical experiment).   

\medskip

\begin{figure}[b!]
\begin{center}

\resizebox{0.79\columnwidth}{!}{%
  \includegraphics{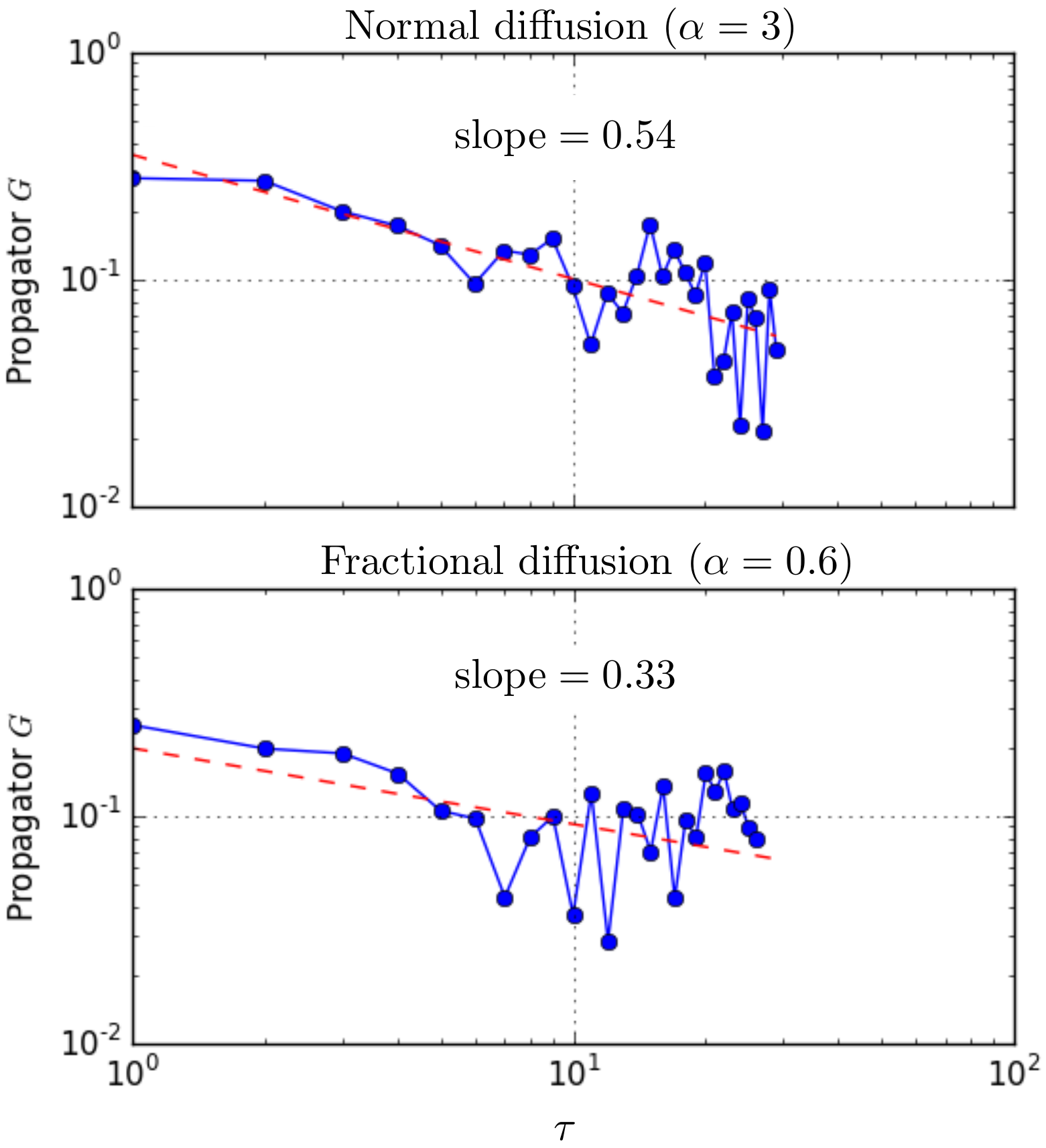}
}
\end{center}
\caption{Fit of the inverted propagator $G_\tau$ for a random order flux simulation: (top) in the normal diffusion case $(\alpha=3>1)$ with slope $0.54\approx1/2$, and (bottom) in the fractional diffusion case $(\alpha=0.6<1)$ with slope $0.33\approx \alpha/2$.}
\label{Propfit}       
\end{figure}

{Using such a simulation to measure impact is not as straightforward as one could think.  In the fast execution limit, discretisation induces the opening of a spread which struggles to refill during the execution. This effect is { naturally} not included in the \emph{continuous} reaction-diffusion model presented above. In particular, it is responsible for the fact that the impact decay is very difficult to reproduce within the numerical framework at hand. Furthermore, the slow execution limit is extremely time consuming, since one must ensure both a small execution rate, meaning $T$ large, while still keeping $T$ small compared to the  cancellation timescale for the above results to hold.\footnote{When $T$ exceeds the cancelation timescale, impact trivially becomes linear in $Q$.}  
\medskip

{ We here use the simulation in order corroborate the central result of the paper that is: fractional diffusion allows to reconciling persistent order flow and diffusive price dynamics (market efficiency).
To do so, }we measure the propagator kernel in the small execution rate limit (see Eq.~\ref{price_eq_slow}). To do so we replace the directional meta-order with a random, IID order flow $m_t=\varepsilon_t$ and follow the corresponding price changes. The discrete linear propagator $G$, defined as $p_t = \sum_{t'<t} G_{t-t'} \varepsilon_{t'}$ \cite{bouchaud2004fluctuations}, can then be obtained by a linear regression and is plotted in Fig.~\ref{Propfit}.  As one can see, in the normal diffusion case $\alpha > 1$ we obtain a square root decaying kernel as predicted by Donier \emph{et al.} \cite{DonierLLOB}, while for the fractional diffusion case $(\alpha<1)$ we obtain a weaker power law decay (with exponent $\approx \alpha/2)$  consistent with Eq.~(\ref{price_eq_slow}).

\section{Conclusion}
\label{sec:6}

We have presented a fractional diffusion extension (FLOB) of the locally linear order book (LLOB) model of Donier \emph{et al.} \cite{DonierLLOB}. The fractional latent order book (FLOB) model presented here is motivated by the existence of a very broad spectrum of time scales in financial markets, spanning seconds to days or even weeks. This extended framework allows us to reconcile the long-memory nature of order flow with market efficiency (\emph{i.e.} diffusive prices). The impact of a metaorder is a concave function of volume, crossing-over from a $Q^{0.75}$ behaviour for small execution rates to the usual $\sqrt{Q}$ behaviour for large execution rates. Another possibility to model agents heterogeneity is to keep a normal diffusion while introducing a wide spectrum of cancellation and deposition rates. Such an approach also yields very interesting, complementary results and is left for an up-coming communication {\cite{ustocome}}.\\

Other possible generalisations can be considered. One is to introduce a broad distribution in the individual volumes of buy/sell orders. This is very relevant in view of the heterogeneity of asset sizes in the financial industry. However, the mathematical apparatus needed to investigate this case needs to be developed, even in standard diffusive case, since fluctuations in the volume profile $\phi(x,t)$ become dominant. Introducing fat-tailed jump lengths (Levy flights) would also be of interest. Indeed, the idea that agents' price reassessments can be large and sudden is actually quite realistic. Again, implementing this idea is not straightforward and will in particular need a careful analysis of the reaction terms, since the continuity of the price paths is lost.\\

In conclusion, we have proposed that the broad distribution of time scales in financial markets could be a crucial ingredient to model the complex dynamics of liquidity and allow one to reproduce realistic price dynamics in stylised Agent-Based Models. \\

We wish to thank  J. Donier (who participated to the first stages of this work), J. Bonart, J. De Lataillade, M. Gould, S. Gualdi, I. Mastromatteo, B. T\'oth and A. ~Darmon for fruitful discussions.


%

%
%
%

%
%

{
\section*{Appendix A: Evolution equation in a CTRW model}
\label{}

We here present the probabilistic derivation of Eq.~\eqref{Ev_eq}. We focus on the case without sources for the sake of simplicity. We denote $P(x,t|x_0,0)$ the conditional probability that a random walker be at position $x$ at time $t$ given that it were at position $x_0$ at time $t=0$. The evolution equation for $P(x,t|x_0,0)$ reads:
\begin{eqnarray}
&&P(x,t|x_0,0)= \Phi(t)\delta(x-x_0) \nonumber\\
&&\quad\quad\quad + \int_0^t \textrm d\tau  \,\Psi(\tau) \hspace{-0.06cm}  \int \textrm d\ell  \Lambda(\ell) P(x,t-\tau|x_0+\ell,0) \ , \quad \quad
\label{probder}
 \end{eqnarray}
 where the first term in the right hand side signifies the particles that were at position $x_0$ at time $t=0$ and have not moved during the interval $[0,t]$, and the second term accounts for the particles which were at position $x_0 + \ell$ at time $t=0$ and have jumped to position $x$ at time $t-\tau$.
 Given the density $\phi(x,0)$ of random walkers at time $t=0$, the particle density $\phi(x,t)$ at position $x$ and time $t$ as:
  \begin{eqnarray}
 \phi(x,t)=\int \textrm d x_0 \phi(x_0,0) P(x,t|x_0,0) \ .
 \end{eqnarray}
 Making use of translation invariance $P(x,t-\tau|x_0+\ell,0)=P(x-\ell,t-\tau|x_0,0)$, multiplying Eq.~\eqref{probder} by $\phi(x_0,0)$, and integrating over $x_0$ yields:
 \begin{equation}
\phi(x,t)= \Phi(t) \phi(x,0) + \int_0^t \textrm d\tau  \hspace{-0.08cm}  \int \textrm d\ell \,\Psi(\tau)   \Lambda(l) \phi(x-\ell,t-\tau) \ .   
\label{phider}
\end{equation}
 Finally, changing variables in Eq.~\eqref{phider} through $\ell \to x-x'$ and $\tau \to t'$ yields Eq.~\eqref{Ev_eq} without sources ($s(x,t)=0$).
  }

\section*{Appendix B: Distribution of waiting times}
\label{}

We here seek a simple analytical form for the Laplace transform of a truncated power law distribution function of waiting times. Having $\Psi(t) \sim \tau_0^\alpha e^{-\epsilon t}/t^{1+\alpha}$, we may compute $\Psi(p)=\int_0^\infty \textrm d t \, e^{-pt} \Psi(t)= 1+ \int_0^\infty \textrm d t ( e^{-pt}-1)\Psi(t)$, where we have made use of the normalisation of $\Psi(t)$. In the limit $t\ll t_c=\epsilon^{-1}$ ($ \epsilon \ll p\ll 1$), one obtains:
\begin{eqnarray}
\Psi(p)-1 \sim \tau_0^\alpha \int_0^\infty \textrm dt\, \frac{e^{-pt}-1}{t^{\alpha+1}} = - (\tau p)^\alpha \ ,
 \end{eqnarray}
where $\tau^\alpha \ := \ - \Gamma[-\alpha] \tau_0^\alpha$ (with  $\Gamma[-\alpha]<0$). In the limit $t\gg t_c=\epsilon^{-1}$ ($ p\ll \epsilon$), one obtains:
\begin{equation}
\Psi(p)-1 \sim  \tau_0^\alpha \int_0^\infty \textrm dt\, (1-pt -1)\frac{e^{-\epsilon t}}{t^{\alpha+1}} = - \Gamma[1-\alpha]\tau_0^\alpha \epsilon^{\alpha - 1 }p \ ,
 \end{equation}
 where $\Gamma[1-\alpha]>0$. The distribution $\Psi(p)$ must thus satisfy $1-\Psi(p) \sim p^{\alpha}$ for $t\ll t_c$ and $1- \Psi(p)\sim p$ for $t\gg t_c$. One can check that the distribution given in Eq.~\eqref{wt_pdf} is indeed an appropriate candidate.

\section*{Appendix C: Fractional diffusion equation}
\label{}

We here present the problem of fractional diffusion at hand in terms of integro-differential equations (see \emph{e.g.}  \cite{Metzler,Schneider,Henry}). Equation~\eqref{solfourier} can also be written in the form:
\begin{equation}
p\phi(k,p) -\phi_0(k)= -  \frac{\omega \,p (k^2+\varphi)}{(p+\epsilon)^\alpha - \epsilon^\alpha}\,\phi(k,p) + s(k,p) \ .
\label{solverseq}
\end{equation}
In the limit $t\ll t_c$, the first term of the right hand side of Eq.~\eqref{solverseq}  reduces to: 
$ - {\omega \,p^{1-\alpha} (k^2+\varphi)}\,\phi(k,p)$. Taking the inverse Fourier-Laplace transform then yields the fractional diffusion equation with death and sources:
\begin{eqnarray}
\partial_t \phi &=& K \mathfrak{D}_t^{1-\alpha}\left( \partial_{xx}\phi -{ \varphi}\phi \right) + s(x,t) \ ,
\label{}
\end{eqnarray}
where $K$ is a generalized diffusion coefficient, and where  $\mathfrak{D}_t^{-\alpha} $ denotes the fractional Riemann-Liouville operator \cite{Schneider,Metzler} defined as  $\mathfrak{D}_t^{-\alpha} f(t) = \Gamma[\alpha]^{-1}\int_0^{t}  {\rm d} {u} \,(t-{u})^{\alpha-1} f({u})$, and $\mathfrak{D}_t^{1-\alpha}=\partial_t \mathfrak{D}_t^{-\alpha}$.\medskip

For $t\gg t_c$, the first term of the right hand side of Eq.~\eqref{solverseq}  reduces to 
$- {\tilde \omega(k^2+\varphi)}\,\phi(k,p)$ and one easily recovers the normal diffusion equation with death and sources $\partial_t \phi = D \partial_{xx}\phi    \ { -\nu\phi} + s(x,t)$, 
where $D= \tilde \omega$ (see \cite{DonierLLOB}).

{
\section*{Appendix D:  Impact in the fast execution limit}
\label{sec:Appendix2}

We here compute the price impact in the limit of large execution rates as given by Eq.~(\ref{largeexecimp}). Letting $x_t = A\sqrt{t}$ into Eq.~(\ref{price_eq}) with a constant execution rate $m_u = m_0$ and changing variables through $u = t(1-v)$ yields:
\begin{equation}
A\sqrt{t} = \frac{t^{1-\alpha/2}}{\mathcal L }  \int_0^1 {\rm d}v \, \frac{m_{{0}}}{\sqrt{4\pi \omega  v^\alpha}}\, g_\alpha \textstyle\Big[\frac{A^2 t^{1-\alpha}}{4\omega  }\frac{\left(1-\sqrt{1-v}\right)^2}{v^\alpha}\Big]   \ . \label{}
\end{equation}
Noting that the integrand dominates for $v\rightarrow 0$ -- which is ${(1-\sqrt{1-v})^2}{v^{-\alpha}} \simeq v^{2-\alpha}/4$ -- and letting $z= vt^{\delta}$ where $\delta = (1-\alpha)/(2-\alpha)$, one obtains:
\begin{equation}
A = \frac{1}{\mathcal L_\alpha }  \int_0^{t^{\delta}} {\rm d}z \, \frac{m_{{0}}}{\sqrt{4\pi \omega  z^\alpha}}\, g_\alpha \left[\frac{A^2 z^{2-\alpha}}{16\omega  }\right]   \, . \label{}
\end{equation}
Letting $w = {A^2 z^{2-\alpha}}/(16\omega  )$ in the limit $A\rightarrow \infty$ yields:
\begin{equation}
A^2 = \frac{2m_0}{\mathcal L_\alpha } \frac1{2-\alpha} \frac1{\sqrt{\pi}}\int_0^{\infty} {\rm d}w \,  \frac{g_\alpha [w]}{\sqrt{w}}   \ , 
\label{}
\end{equation}
and making use of the normalisation of the fractional diffusion kernel $\int {\rm d}x \,\mathcal G_{\alpha}(x,t)= 1 $, which together with Eq.~(\ref{Gg}) can also be written as $\int_0^{\infty} {\rm d}w \,  {g_\alpha [w]}/{\sqrt{w}} = \sqrt{\pi}$, finally yields Eq.~(\ref{largeexecimp}).

}

\bibliographystyle{iopart-num}
\bibliography{bibs}

\end{document}